\documentclass[aps,prd,twocolumn,floatfix,showpacs]{revtex4}

\usepackage{graphicx} 
\usepackage{dcolumn} 
\usepackage{bm} 

\def\met{\mbox{${\hbox{$E$\kern-0.6em\lower-.1ex\hbox{/}}}_T$}} 
\def\lsim{\mathrel{\rlap{\lower4pt\hbox{\hskip1pt$\sim$}}
    \raise1pt\hbox{$<$}}}         
\def\gsim{\mathrel{\rlap{\lower4pt\hbox{\hskip1pt$\sim$}}
    \raise1pt\hbox{$>$}}}         

\begin{document}
\bibliographystyle{apsrev}

\hspace{5.2in} \mbox{Fermilab-Pub-07-493-E}


\title{Measurement of the muon charge asymmetry from $\bm{W}$ boson decays}

%
\author{V.M.~Abazov$^{35}$}
\author{B.~Abbott$^{75}$}
\author{M.~Abolins$^{65}$}
\author{B.S.~Acharya$^{28}$}
\author{M.~Adams$^{51}$}
\author{T.~Adams$^{49}$}
\author{E.~Aguilo$^{5}$}
\author{S.H.~Ahn$^{30}$}
\author{M.~Ahsan$^{59}$}
\author{G.D.~Alexeev$^{35}$}
% ...

\affiliation{\vspace{0.1 in}(The D\O\ Collaboration)\vspace{0.1 in}}
% ...
\affiliation{$^{5}$University of Alberta, Edmonton, Alberta, Canada,
                Simon Fraser University, Burnaby, British Columbia, Canada,
                York University, Toronto, Ontario, Canada, and
                McGill University, Montreal, Quebec, Canada}
% ...
\affiliation{$^{28}$Tata Institute of Fundamental Research, Mumbai, India}
% ...
\affiliation{$^{30}$Korea Detector Laboratory, Korea University, Seoul, Korea}
% ...
\affiliation{$^{35}$Joint Institute for Nuclear Research, Dubna, Russia}
% ...
\affiliation{$^{49}$Florida State University, Tallahassee, Florida 32306, USA}
% ...
\affiliation{$^{51}$University of Illinois at Chicago,
                Chicago, Illinois 60607, USA}
% ...
\affiliation{$^{59}$Kansas State University, Manhattan, Kansas 66506, USA}
% ...
\affiliation{$^{65}$Michigan State University,
                East Lansing, Michigan 48824, USA}
% ...
\affiliation{$^{75}$University of Oklahoma, Norman, Oklahoma 73019, USA}
% ...

\date{September 25, 2007}

\begin{abstract}
We present a measurement of the muon charge asymmetry from $W$ boson decays 
using 0.3~fb$^{-1}$
of data collected
at $\sqrt{s}=1.96$~GeV between 2002 and 2004 with the D0 detector at the 
Fermilab Tevatron $p\overline{p}$ Collider.  We compare our findings with 
expectations from next-to-leading-order calculations performed using the 
CTEQ6.1M and MRST04 NLO parton distribution functions.  Our findings 
can be used to constrain future parton distribution function fits.
\end{abstract}

\pacs{13.38.Be,13.85.Qk,14.60.Ef,14.70.Fm}

\maketitle

A measurement of the $W^\pm$ boson rapidity ($y_W$) distributions in
$p\overline{p}$ collisions provides valuable information about the parton
distribution functions (PDFs) of the $u$ and $d$ quarks in the proton.  
$W$ bosons at the Fermilab Tevatron $p\overline{p}$ Collider are primarily 
produced by quark-antiquark annihilation.  Contributions from valence-valence 
and valence-sea annihilations provide about 85\% of the cross section 
with the rest coming from
sea-sea annihilations.  A $W^+$ boson is usually produced by the
interaction between a $u$ quark from the proton and a $\overline{d}$ quark from
the antiproton, while a $W^-$ boson is predominantly produced by a $d$ quark 
from the proton and a $\overline{u}$ quark from the antiproton.  Since $u$ 
valence quarks
carry on average more of the momentum of the proton than $d$ valence quarks
\cite{u-more1,u-more2}, $W^+$ bosons tend to move in the proton direction
and $W^-$ bosons in the antiproton direction, giving rise to the $W$ boson
production charge asymmetry.

The $W^\pm$ boson production asymmetry provides complementary information on
the PDFs to that 
from deep inelastic scattering experiments.  In particular, it contributes to
determining the slope of the $d/u$ quark ratio in the region $x\lsim 0.3$, 
where $x$ is the momentum fraction carried by a parton in the proton 
\cite{CTEQ3,CTEQ4}.
Knowledge of the $d/u$ ratio is needed for predicting the transverse 
momentum ($p_T$) spectrum of leptons from $W$ boson decay, an important
ingredient in the precision measurement of the $W$ boson mass.

The region of phase space in $x$ that can be probed depends on the range of 
the rapidity of the $W$ boson and, at leading order, is given by 
\begin{equation}
x_{1(2)} = \frac{M_W \cdot c^2}{\sqrt{s}} e^{(\pm)y_W},
\end{equation}
where $x_{1(2)}$ is the momentum fraction carried by the $u$($d$) quark,
$+(-)y_W$ is the rapidity of the positive (negative) $W$ boson, and 
$\sqrt{s}$ is the
center of mass energy.  For this analysis, $-2 \lsim y_W \lsim 2$ and 
$\sqrt{s} = 1.96$\ GeV, allowing us to probe $0.005 \lsim x \lsim 0.3$.

The $W$ boson rapidity is not directly measurable since the longitudinal 
momentum
of the neutrino from its decay cannot be determined.  However, the rapidity 
distribution of the charged lepton from the $W$ boson decay reflects this 
asymmetry; here we use muons for this purpose.  The muon asymmetry
is a convolution of the $W$ boson production asymmetry and the asymmetry from
the $(V-A)$ nature of the $W$ boson decay.  Since the $(V-A)$ interaction is
well understood, the muon charge asymmetry can be used to probe the PDFs.

The muon charge asymmetry as a function of the muon rapidity is defined as
\begin{equation}
A(y) = \frac{ \frac{d\sigma}{dy}(\mu^+) - \frac{d\sigma}{dy}(\mu^-) }
{\frac{d\sigma}{dy}(\mu^+) + \frac{d\sigma}{dy}(\mu^-) }
\end{equation}
where $\frac{d\sigma}{dy}(\mu^\pm)$ is the differential 
cross section for the $W^\pm$ boson decay muons as a function of muon rapidity 
$y$.  In this analysis, we measure the
muon charge asymmetry as a function of pseudorapidity 
$\eta = -\ln[\tan({\theta/2})]$,
where $\theta$ is the polar angle with respect to the proton beam.  

Allowing for acceptance and efficiency differences between positively and
negatively charged muons, the muon charge asymmetry can be written as 
\begin{equation}
\label{eq:muasym}
A(\eta) = \frac{ N_{\mu^+}(\eta) - k(\eta)N_{\mu^-}(\eta)}
{N_{\mu^+}(\eta) + k(\eta)N_{\mu^-}(\eta)}
\end{equation}
where $\eta$ is the pseudorapidity of the muon, 
$N_{\mu^\pm}(\eta)$ is the background-corrected number of muons in
pseudorapidity bin $\eta$, 
and $k(\eta)=\epsilon^+(\eta)/\epsilon^-(\eta)$ is the 
acceptance and efficiency ratio between the positive and negative muons.

The lepton charge asymmetry from $W$ boson decay was measured by the CDF
collaboration in the electron and muon channels during Run I of the 
Tevatron Collider 
\cite{cdf1,cdf2} and in the electron channel using Run II data \cite{cdf3}.  
The measurement described here is based on a larger data sample than these
analyses used.  In addition, the muon channel benefits from a much lower charge 
misidentification probability than the electron channel has.

For this analysis, we used 0.3~fb$^{-1}$
of data collected using
the D0 detector at a center-of-mass energy of 1.96~TeV at the Tevatron 
Collider.  The D0 detector is described in detail in
Ref.~\cite{run2-nim}; here we provide a brief description.  
The detector consists of a central tracking system, 
calorimeters, and a muon detector.  The central tracking system comprises a
silicon microstrip tracker (SMT) and a central fiber tracker (CFT), both located
within a 2~T solenoidal magnet.  The SMT has a six-barrel longitudinal structure
interspersed with sixteen radial disks.  The CFT has eight coaxial barrels, each
supporting two doublets of overlapping scintillating fibers, measuring along the
axial direction and at stereo angles of $\pm3^\circ$.  The liquid-argon/uranium
calorimeter is in three sections, a central section covering $|\eta|\lsim 1$ and
two end caps extending coverage to $|\eta|\approx 4$.  The calorimeter is
surrounded by the muon detector which consists of three layers of scintillators
and drift tubes, one layer in front of a 1.8~T iron toroidal magnet and two 
layers 
outside the magnet.  Coverage for muons is partially compromised by the 
calorimeter support structure at the bottom of the detector ($|\eta|<1.25$ and 
$4.25<\phi<5.15$, where $\phi$ is the azimuthal angle).

$W\to\mu\nu$ events were collected using two single-muon triggers.  An
unprescaled trigger, the ``wide'' trigger, covered the region $|\eta|<1.5$, 
and a prescaled trigger, the ``all'' trigger, covered $|\eta|<2$.  Both 
triggers required hits in the muon system consistent in location and timing with
a muon originating from the interaction region at the first trigger level, a 
reconstructed track in the muon system with 
$p_T > 3$\ GeV/$c$ at the second trigger level, and a track in the central 
tracking system with $p_T > 10$\ GeV/$c$ at the third trigger level.  Offline, 
muon candidates were required to lie within the geometrical acceptance of the 
detector, to have reconstructed track segments in at least two layers of the
muon system, and to be matched to 
a track in the central tracking system with transverse momentum 
$p_T>20$~GeV/$c$.  To ensure well-measured tracks, the matching track was 
required to have at least one hit in the SMT, at least nine hits in the CFT, 
a track fit $\chi^2/{\text{dof}} < 3.3$, and a distance of closest approach to 
the event vertex $|{\text{dca}}| < 0.011$\ cm.  Muon candidates were also 
required to be isolated in both the central tracking system and the calorimeter:
the sum of the transverse momenta of the tracks in a cone of radius 0.5 in 
$\eta-\phi$ space 
around the muon track had to be less than 2.5 GeV/$c$, and the total transverse 
energy in a hollow cone of inner radius 0.1 and outer radius 0.4 around the
extrapolated muon position in the calorimeter had to be less than 2.5 GeV.
Events with $|\eta|<1.4$ ($1.4<|\eta|<2$) were required to satisfy the 
wide (all) trigger.  $W$ boson candidates were further selected by
requiring the missing transverse energy $\met$, determined by the vector sum 
of the transverse components of the energy deposited in the calorimeter and the 
$p_T$ of the muon, to be greater than 20 GeV and the transverse 
mass $M_T > 40$\ GeV/$c^2$.  The analysis was done in muon pseudorapidity bins 
of width 0.2; each bin was treated independently.  The pseudorapidity
resolution of the tracking system is approximately 0.01 for $|\eta| < 1.7$ and
is not expected to worsen significantly up to $|\eta| = 2$.

Events characteristic of the $Z\to\mu\mu$ background were removed using two
criteria:  (i) all events with a second muon were rejected, except those 
within $|\Delta\phi| < 0.1$ of the selected muon to avoid vetoing events
containing multiple muons reconstructed from a single real muon at detector 
boundaries, and 
(ii) all events containing a second track 
with $|\Delta\phi| > 2.1$ from the selected muon, $p_T> 20$\ GeV/$c$, and 
$|{\text{dca}}| < 0.011$\ cm were rejected.  Cosmic ray muons
were rejected using muon system scintillator timing cuts and by the dca 
requirement.  A total of 189697 $W\to\mu\nu$ candidates was selected.

The asymmetry measurement is sensitive to the misidentification of the charge of
the muon.  A positive muon misidentified as a negative muon would not only add
to the number of negative muons, but would also reduce the number of
positive muons, and vice versa, diluting the true asymmetry.  The charge
misidentification probability was estimated using a dimuon data sample in which 
events were required to satisfy one of the two single-muon triggers described 
above.  Events containing two muons satisfying all of the above conditions
(except the second muon veto) and 
with a dimuon invariant mass $M_{\mu\mu} > 40$\ GeV/$c^2$ were selected to form 
a $Z$ boson sample.  The event sample contained 9958 events 
and only one same-sign dimuon event, giving an average charge 
misidentification probability of 
$(5.0^{+12}_{-4.2}) \times 10^{-5}$
over $ -2 < \eta < 2$.  Removing
the dimuon invariant mass cut or lowering the muon $p_T$ cut to 15 GeV/$c$ did 
not change the result.  The probability was verified using an
independently-triggered
dimuon sample in which events were collected using one of a set of dimuon 
triggers with no track requirements. 
Out of 19284 dimuon events selected as described above, two had 
same-sign muons.  We also measured the probability 
using a sample of $W\to\mu\nu$ events generated with {\sc{pythia}} v6.2 
\cite{pythia62} and CTEQ6.1M PDFs \cite{CTEQ1} and passed through the full D0
detector simulation based on {\sc{geant}} \cite{geant}; it 
is  approximately the same as that determined using the data.  
Therefore, charge misidentification is not expected to 
influence the final asymmetry measurement.  To determine the systematic
uncertainty on the asymmetry due to charge misidentification, we used Poisson
uncertainties based on the number of muon tracks in each $\eta$ bin from the
single-muon-triggered $Z$ boson sample and varied the asymmetry accordingly.  
The largest uncertainty due to charge misidentification is in the range 
$1.8 < \eta \leq 2.0$, where it is 0.001.

Ideally, the acceptances and efficiencies would be the same for all muons as a
function of $\eta$ and independent of charge and $p_T$, leading to $k(\eta) =
1$.  To reduce charge effects due to detector asymmetries, the directions of the
magnetic fields in the solenoidal and toroidal magnets were regularly reversed.
Approximately 51.1\% of the selected $W$ boson sample was collected with the 
solenoid at forward polarity, with 48.9\% at reverse polarity.  For the toroid, 
50.7\% (49.3\%) of the selected sample was collected with forward (reverse)
toroid polarity, respectively.  In addition, we studied the trigger efficiency 
for muons at the first two
trigger levels, the trigger efficiency for tracks at the third trigger level,
the offline muon reconstruction efficiency, the offline tracking
efficiency, and the isolation efficiency.  The four trigger and reconstruction
efficiencies are discussed together while the isolation efficiency is
discussed separately in the following text.

\begin{figure}
\includegraphics[width=7.8cm]{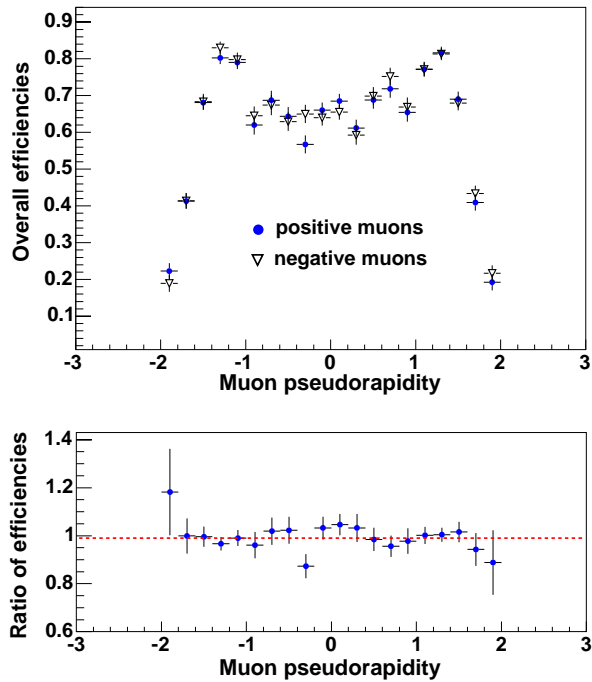}
\caption{(a) The combined trigger and reconstruction efficiency distributions 
by charge as a function of $\eta$ and (b) the ratio of these efficiencies fit 
to a constant, $0.99\pm 0.01$.}
\label{fig:comb-eff}
\end{figure}

All efficiencies were measured using the tag and probe method on a sample of
dimuon events collected using one of the single-muon triggers.  To select 
primarily $Z$
boson decays, events were required to have $M_{\mu\mu} > 40$\ GeV/$c^2$.  
First a tag muon was chosen as a track-matched, isolated muon 
satisfying all of the 
selection conditions.  The probe was another track or muon whose selection 
criteria depended on the efficiency being studied.  All efficiencies were 
checked as functions of $p_T$, charge, and $\eta$.  No dependence 
on $p_T$ or charge was observed for the four trigger and reconstruction
efficiencies.  The four efficiencies were multiplied to 
determine the combined efficiencies for positive and negative muons as 
functions of $\eta$.  Figure \ref{fig:comb-eff} shows the combined efficiencies 
by charge and the ratio of these efficiencies as functions of $\eta$.  The 
ratio was fit to a constant value of $0.99 \pm 0.01$, with a 
$\chi^2/{\text{dof}}$ of 0.71.

\begin{figure}
\includegraphics[width=7.8cm]{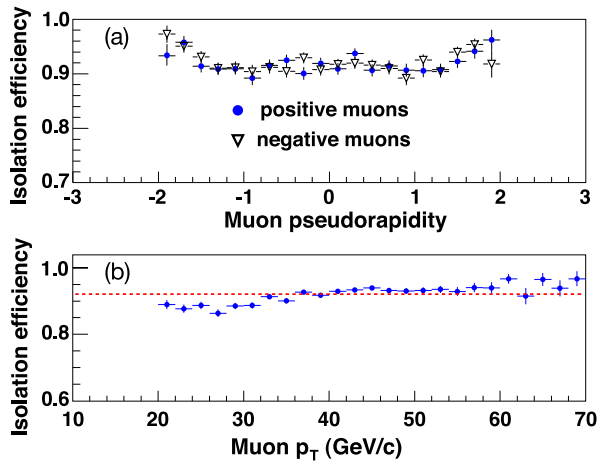}
\caption{The isolation efficiency as a function of (a) $\eta$ (separated by
muon charge) and (b) $p_T$ (for all muons) fit to a constant 
$0.921 \pm 0.002$.}
\label{fig:iso-eff}
\end{figure}

The isolation efficiency was also measured using the tag and probe method on 
a sample of dimuon events.  One muon candidate was required to satisfy all muon
and track selection requirements, while the other had to satisfy all
requirements except that it was not required to be isolated in either the
tracker or the calorimeter.  The fraction of isolated probe tracks with
$M_{\mu\mu} > 40$~GeV/$c^2$ gives the efficiency.  The isolation efficiency is 
shown in Fig.\ \ref{fig:iso-eff} as a function of $p_T$ and of $\eta$.  The 
efficiency is constant for, and consistent between, both charges 
over the full $\eta$ range.  The efficiency as a function of $p_T$ shows a 
slight $p_T$ dependence.  We chose, however, to fit the efficiency to a 
constant value of $0.921 \pm 0.002$ with a $\chi^2/{\text{dof}} = 5.8$.  To 
account for the high $\chi^2/{\text{dof}}$, the uncertainty was determined 
from the 
isolation efficiency distribution itself; the rms of this distribution is 
0.022, and this value was used as the isolation efficiency uncertainty.

The largest source of contamination in the data sample comes from electroweak
processes,  $Z \to \mu\mu$ and $W\to\tau\nu$ and 
$Z\to\tau\tau$ where $\tau\to\mu\nu\nu$.  Muons from these electroweak decays 
exhibit charge asymmetries which dilute the true asymmetry.  These asymmetries 
are accounted for by subtracting the background bin-by-bin in $\eta$.
The electroweak background was estimated using Monte Carlo samples generated 
with {\sc pythia} v6.2 and CTEQ6.1M PDFs and a 
parameterized description of the D0 
detector \cite{hesketh}.  For each of the three processes, separate samples 
were generated for the two triggers.  

The background due to semi-leptonic decays and punch-through in multijet 
events was estimated using the data.  The isolation criteria remove
events containing muons within jets, and they were used as the discriminator to
determine this background.  Using a sample of events passing all selection
criteria except that on the transverse mass and requiring $\met < 10$ GeV, we 
measured the probability for multijet events to satisfy the isolation criteria.
This probability shows no dependence on muon pseudorapidity.

Table~\ref{tab:backgrounds} shows the overall contribution of each of the four
backgrounds.  To determine the number of events for each background, the
contributions from the three 
electroweak processes were added to the number of events expected from 
$W\to\mu\nu$ Monte Carlo events produced using {\sc pythia} 
and the parameterized detector description, and this sum was normalized to the 
number of events in the data less the estimated multijet background.  The 
overall normalization was done for $|\eta|<1.5$ for the wide trigger and 
$|\eta|<2$ for the all trigger, while for the final result, it was done
independently in each $\eta$ bin.  This is the only use of signal Monte Carlo
events in the analysis.

\begin{table}
\caption{Estimated backgrounds in the $W$ boson sample by trigger. 
Uncertainties are statistical only.}
\begin{ruledtabular}
\begin{tabular}{ccc}
Background & Wide trigger & All trigger \\  \hline
$Z\to\mu\mu$   & (4.31 $\pm$ 0.05)\% & (4.39 $\pm$ 0.11)\% \\ 
$Z\to\tau\tau$ & (0.19 $\pm$ 0.01)\% & (0.20 $\pm$ 0.02)\% \\ 
$W\to\tau\nu$  & (2.32 $\pm$ 0.02)\% & (2.43 $\pm$ 0.08)\% \\ 
Multijet events & (2.77 $\pm$ 0.04)\% & (2.76 $\pm$ 0.09)\% \\ 
\end{tabular}
\end{ruledtabular}
\label{tab:backgrounds}
\end{table}

\begin{figure}
\includegraphics[width=8.2cm]{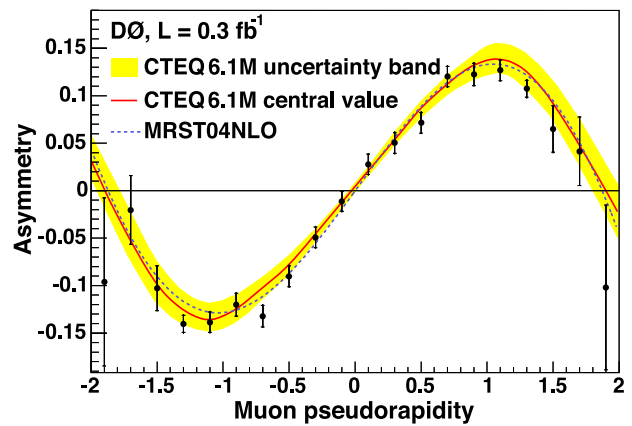}
\caption{The muon charge asymmetry distribution.  
The horizontal bars show the statistical uncertainty and the full vertical
lines show the total uncertainty on each point. The shaded (yellow) band is 
the envelope determined using the forty CTEQ6.1M PDF uncertainty sets, the 
solid (red) line is the CTEQ6.1M central value, and the dotted (blue) line is 
the charge asymmetry determined using the MRST04 NLO PDFs.  All three were
determined using {\sc resbos} and {\sc photos} (color online).}
\label{fig:asym}
\end{figure}

The muon charge asymmetry was determined separately for each bin in $\eta$ and
is shown in Fig.\ \ref{fig:asym}.  Also shown are the asymmetry determined 
using the {\sc resbos} \cite{resbos} event generator with QCD resummation 
and {\sc photos} \cite{photos} NLO QED corrections in the final state with
the CTEQ6.1M PDFs, with the 
forty CTEQ6.1M  PDF uncertainty sets \cite{CTEQ1,CTEQ2}, and with the MRST04
NLO PDFs \cite{MRST}.  The next-to-next-to-leading-order calculation of the $W$ 
boson rapidity distribution \cite{nnlo-rapidity} at the Tevatron is very similar
to the next-to-leading-order calculation.  The $W$ boson asymmetry distribution
is not very sensitive to QCD corrections, and calculations at \hbox{leading-,}
next-to-leading-, and next-to-next-to-leading-order are nearly indistinguishable
\cite{nnlo-rapidity}.

Systematic uncertainties taken into account are those on the ratio of the 
efficiencies for positively and negatively charged muons [$k(\eta)$], the 
efficiency of the isolation criteria, the charge misidentification, the 
probability for multijet events to satisfy the isolation criteria, the 
hadronic energy scale for the detector (needed to calculate \met), 
and the parameterization of the muon 
energy loss in the calorimeter.  Each contribution was varied by $\pm 1\sigma$, 
and the asymmetry was recalculated.  The changes in the measured values were 
added in quadrature to determine the overall systematic uncertainty.  
The measured asymmetry and the statistical and systematic 
uncertainties are listed in Table \ref{tab:asym_err} for each $\eta$ bin.  
The 
efficiency ratio uncertainty, which is based on the number of dimuon events in 
the single-muon-triggered samples, dominates and is
approximately equal to 0.005 in all $\eta$ bins.

\begin{table}
\caption{The measured muon asymmetry in bins of pseudorapidity, calculated 
using Eq.\ \ref{eq:muasym}.  The asymmetry values are the averages within each
pseudorapidity bin.  The first 
uncertainty is statistical; the second is systematic.}
\begin{ruledtabular}
\begin{tabular}{D{,}{\,-\,}{-1} D{,}{\,\pm\,}{-1}}
\multicolumn{1}{r}{{\phantom{xxxx}}Pseudorapidity range}  & 
\multicolumn{1}{r}{Muon asymmetry{\phantom{xx}}} 
\\ \hline
 -2.0 , -1.8   & -0.096  , 0.089 \pm 0.005  \\
 -1.8 , -1.6   & -0.020  , 0.036 \pm 0.005  \\
 -1.6 , -1.4   & -0.103  , 0.024 \pm 0.005  \\
 -1.4 , -1.2   & -0.140  , 0.009 \pm 0.005 \\
 -1.2 , -1.0   & -0.138  , 0.011 \pm 0.005  \\
 -1.0 , -0.8   & -0.120  , 0.012 \pm 0.005  \\
 -0.8 , -0.6   & -0.132  , 0.011 \pm 0.005  \\
 -0.6 , -0.4   & -0.090  , 0.011 \pm 0.005  \\
 -0.4 , -0.2   & -0.049  , 0.011 \pm 0.005  \\
 -0.2 ,  0.0   & -0.011  , 0.010 \pm 0.005  \\
  0.0 ,  0.2   &  0.028  , 0.011 \pm 0.005  \\
  0.2 ,  0.4   &  0.050  , 0.011 \pm 0.005  \\
  0.4 ,  0.6   &  0.071  , 0.011 \pm 0.005  \\
  0.6 ,  0.8   &  0.120  , 0.011 \pm 0.005  \\
  0.8 ,  1.0   &  0.122  , 0.012 \pm 0.005  \\
  1.0 ,  1.2   &  0.127  , 0.011 \pm 0.005  \\
  1.2 ,  1.4   &  0.107  , 0.009 \pm 0.005  \\
  1.4 ,  1.6   &  0.065  , 0.025 \pm 0.007  \\
  1.6 ,  1.8   &  0.042  , 0.036 \pm 0.005  \\
  1.8 ,  2.0   & -0.102  , 0.087 \pm 0.005  \\
\end{tabular}
\end{ruledtabular}
\label{tab:asym_err}
\end{table}

\begin{figure}
\includegraphics[width=8.2cm]{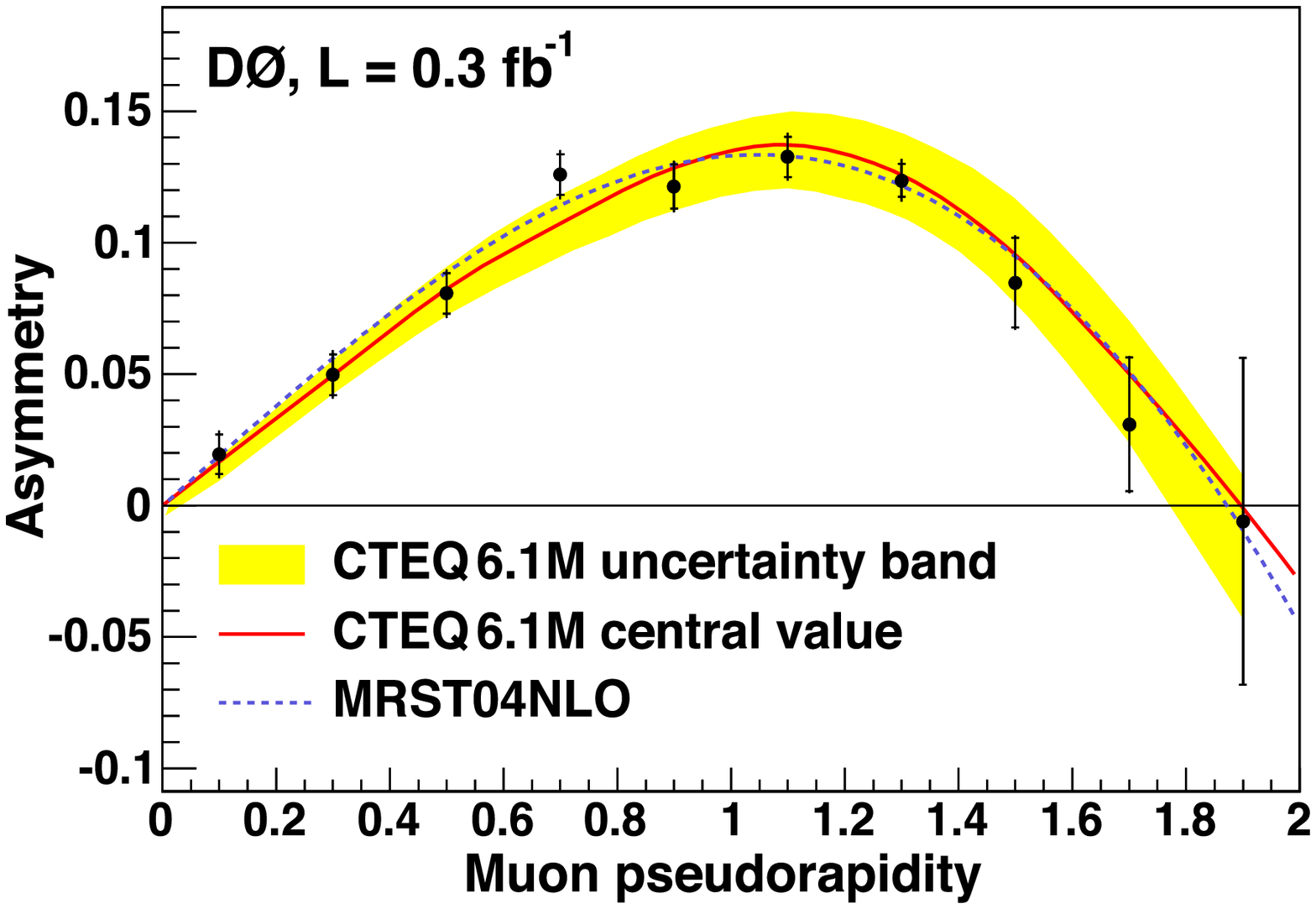}
\caption{The folded muon charge asymmetry distribution.  
The horizonal bars show the statistical uncertainty and the full vertical
lines show the total uncertainty on each point. The shaded (yellow) band is 
the envelope determined using the forty CTEQ6.1M PDF uncertainty sets, the 
solid (red) line is the CTEQ6.1M central value, and the dotted (blue) line is 
the charge asymmetry determined using the MRST04 NLO PDFs.  All three were
determined using {\sc resbos} and {\sc photos} (color online).}
\label{fig:folded}
\end{figure}

By CP invariance, the asymmetries at $\pm\eta$ have opposite signs and equal
magnitudes, allowing the asymmetry distribution to be ``folded'' to decrease 
the statistical uncertainty.  The folded asymmetry distribution was found 
by combining the numbers of events in each $\pm\eta$ bin and redetermining the
background and systematic uncertainties as described above.  The folded 
distribution is shown in Fig.\ \ref{fig:folded} with the measured values of 
the asymmetry and uncertainties given in Table \ref{tab:folded}.  For 
$0.7 \lsim |\eta| \lsim 1.3$, our experimental uncertainties are smaller than
the uncertainty given by the CTEQ uncertainty sets.  Only at the extremes of 
our measurement, in the 0.0--0.2 and 1.8--2.0 muon pseudorapidity bins, are our
uncertainties larger than the CTEQ uncertainty.  Between these region, the
uncertainties are comparable.

\begin{table}
\caption{The folded muon asymmetry in bins of pseudorapidity. 
The asymmetry values are the averages within each pseudorapidity bin. 
The first uncertainty is statistical; the second is systematic.}
\begin{ruledtabular}
\begin{tabular}{D{,}{\,-\,}{-1} D{,}{\,\pm\,}{-1}}
\multicolumn{1}{r}{{\phantom{xx}}Pseudorapidity range}  & 
\multicolumn{1}{r}{Muon asymmetry{\phantom{xx}}} 
\\ \hline
  0.0 , 0.2  &  0.019 , 0.008 \pm 0.005 \\
  0.2 , 0.4  &  0.050 , 0.008 \pm 0.005 \\
  0.4 , 0.6  &  0.081 , 0.008 \pm 0.005 \\
  0.6 , 0.8  &  0.126 , 0.008 \pm 0.005 \\
  0.8 , 1.0  &  0.121 , 0.008 \pm 0.005 \\
  1.0 , 1.2  &  0.133 , 0.008 \pm 0.005 \\
  1.2 , 1.4  &  0.124 , 0.006 \pm 0.005 \\
  1.4 , 1.6  &  0.085 , 0.017 \pm 0.006 \\
  1.6 , 1.8  &  0.031 , 0.026 \pm 0.005 \\
  1.8 , 2.0  & -0.006 , 0.062 \pm 0.005 \\
\end{tabular}         
\end{ruledtabular}
\label{tab:folded}
\end{table}

We have measured the charge asymmetry of muons from $W$ boson decay using
0.3~fb$^{-1}$ of data.  Our results can already improve
constraints on the PDFs.  In the future, measurement of the $W$ boson 
asymmetry will have a significant impact on PDF determination as present 
uncertainties are dominated by statistics.

We thank the staffs at Fermilab and collaborating institutions, 
and acknowledge support from the 
DOE and NSF (USA);
CEA and CNRS/IN2P3 (France);
FASI, Rosatom and RFBR (Russia);
CAPES, CNPq, FAPERJ, FAPESP and FUNDUNESP (Brazil);
DAE and DST (India);
Colciencias (Colombia);
CONACyT (Mexico);
KRF and KOSEF (Korea);
CONICET and UBACyT (Argentina);
FOM (The Netherlands);
Science and Technology Facilities Council (United Kingdom);
MSMT and GACR (Czech Republic);
CRC Program, CFI, NSERC and WestGrid Project (Canada);
BMBF and DFG (Germany);
SFI (Ireland);
The Swedish Research Council (Sweden);
CAS and CNSF (China);
Alexander von Humboldt Foundation;
and the Marie Curie Program.

\vskip0.3cm


\noindent 
{[a]} Visitor from Augustana College, Sioux Falls, SD, USA. \\
{[b]} Visitor from The University of Liverpool, Liverpool, UK. \\
{[c]} Visitor from ICN-UNAM, Mexico City, Mexico. \\
{[d]} Visitor from II.\ Physikalisches Institut,
Georg-August-University G{\"o}ttingen, Germany. \\
{[e]} Visitor from Helsinki Institute of Physics, Helsinki, Finland. \\
{[f]} Visitor from Universit{\"a}t Z{\"u}rich, Z{\"u}rich, Switzerland. \\
{[\dag]} Fermilab International Fellow. \\
{[\ddag]} Deceased.

\bibliography{Wasym-prd}

\end{document}